\documentclass[conference]{IEEEtran}
\IEEEoverridecommandlockouts
\usepackage{amsmath,amssymb,amsfonts}
\usepackage{graphicx}
\usepackage{textcomp}

\PassOptionsToPackage{disable}{todonotes}
\PassOptionsToPackage{final}{showkeys}

\PassOptionsToPackage{frozencache,cachedir=.}{minted}

\usepackage{garyw}

\hypersetup{
colorlinks=true,
allcolors=RoyalBlue
}

\usepackage[
numbers,square
]{natbib}

\newcommand{\oto}{One-to-one}
\newcommand{\ota}{One-to-all}
\newcommand{\ooto}{Opt-one-to-one}

\newcommand{\colismall}{E. Coli 29X}
\newcommand{\colilarge}{E. Coli 100X}

\begin{document}

\title{
GPU Scheduler for \textit{De Novo} Genome Assembly with Multiple MPI Processes
\thanks{
\textsuperscript{*} All authors have contributed equally.
Project advised by Prof. Giulia Guidi (\href{mailto:gg434@cs.cornell.edu}{gg434@cs.cornell.edu}) of Cornell University.
}
}

\author{
\IEEEauthorblockN{Minhao Li*}
\IEEEauthorblockA{\textit{Department of Computer Science} \\
\textit{Cornell University}\\
Ithaca, NY 14853 \\
\texttt{ml2499@cornell.edu}}
\and
\IEEEauthorblockN{Siyu Wang*}
\IEEEauthorblockA{\textit{Department of Computer Science} \\
\textit{Cornell University}\\
Ithaca, NY 14853 \\
\texttt{sw988@cornell.edu}}
\and
\IEEEauthorblockN{Guanghao Wei*}
\IEEEauthorblockA{\textit{Department of Computer Science} \\
\textit{Cornell University}\\
Ithaca, NY 14853 \\
\texttt{gw338@cornell.edu}}
}

\maketitle

\begin{abstract}
\textit{De Novo} Genome assembly is one of the most important tasks in computational biology. 
ELBA is the state-of-the-art distributed-memory parallel algorithm for overlap detection and layout simplification steps of \textit{De Novo} genome assembly but exists a performance bottleneck in pairwise alignment.

In this work, we proposed 3 GPU schedulers for ELBA to accommodate multiple MPI processes and multiple GPUs. 
The GPU schedulers enable multiple MPI processes to perform computation on GPUs in a round-robin fashion. 
Both strong and weak scaling experiments show that 3 schedulers are able to significantly improve the performance of baseline while there is a trade-off between parallelism and GPU scheduler overhead.
For the best performance implementation, the one-to-one scheduler achieves $\sim$7-8$\times$ speed-up using 25 MPI processes compared with the baseline vanilla ELBA GPU scheduler.
\end{abstract}

\section{Introduction}
\label{sec:intro}

\textit{De Novo} Genome assembly is the method for constructing genomes from a large number of (short- or long-) DNA fragments, with no prior knowledge of the correct sequence or order of those fragments.
It is one of the most important and computationally intensive tasks in computational biology.

ELBA~\cite{Guidi2020ParallelSG} is the state-of-the-art distributed-memory parallel algorithm that uses MPI, OpenMP, and sparse matrix multiplication to accelerate the task on the CPU. 
However, \citet{Zeni2020LOGANHG} shows that pairwise alignment constitutes about 90\% of the overall run-time when using real data sets of genome assembly tasks.
Additionally, \citet{Zeni2020LOGANHG} shows that taking advantage of high-performance computing using GPU and CUDA boosts the performance of pairwise alignment.

Our work proposes 3 schemes of GPU scheduler, namely \oto{}, \ota{}, and \ooto{}.
All 3 schedulers show great strong scaling efficiency.
For the best performance implementation, the \oto{} scheduler achieves $\sim$7-8$\times$ speed-up using 25 MPI processes compared with the baseline vanilla ELBA GPU scheduler.

For the rest of this paper, \cref{sec:method} introduces the methodology and scheme of 3 schedulers, \cref{sec:results} presents the empirical results of our experiments and evaluates the pros and cons of these schedulers.

The implementation is highly based on the code base of ELBA\footnote{\url{https://github.com/PASSIONLab/ELBA}}, and our implementation is publicly available on GitHub\footnote{\url{https://github.com/garywei944/ELBA/tree/GPU}}.

\section{Related works}
\label{sec:related}

In this section, we reviewed the prior work of ELBA in the literature that led to the proposal of this work.

BELLA~\cite{Guidi2018BELLABE} is the first work that formulates overlap detection for \textit{De Novo} genome assembly using sparse matrices under a distributed memory setting.
BELLA uses a seed-based approach to detect overlaps in the context of long-read applications and uses a Markov chain model to filter out unreliable k-mers. 
If the sparse overlap matrix is too large, BELLA divides it into batches based on available RAM.
For the alignment phase, BELLA proposes an efficient seed-and-extend algorithm.
An adaptive threshold is used to perform the X-drop alignment.
Experiments show that using the probability model, BELLA improves the assembly quality.
It also achieves a 2.5$\times$ boost in performance compared to BLASR~\cite{Chaisson2012MappingSM}.

As a follow-up work of BELLA, diBELLA 2D~\cite{Guidi2020ParallelSG} uses the Overlap-Layout-Consensus paradigm to tackle the \textit{De Novo} genome assembly problem, which introduces the theoretical background of ELBA.
The algorithm uses a sparse linear algebra-centric approach to optimize the overlap and layout phases. One important component is distributed Sparse General Matrix Multiply, which has high parallelism and is used to perform and boost the overlap step.
The input data are divided into equal-sized independent chunks for MPI processes. 
For the pairwise alignment, diBELLA 2D uses an algorithm similar to X-Drop.
Experiments show that the overall pipeline has great strong scaling with respect to the number of MPI processes.
The most time-consuming phase is pairwise alignment, which takes approximately more than 90\% of the overall time, which provides us with a starting point and encourages us to keep the MPI parallelism.

LOGAN is a near-optimal implementation of the X-Drop, a dynamic programming algorithm that solves pairwise alignment, using CUDA to boost the performance of pairwise alignment tasks.
Multiple layers of parallelism including intra-sequence parallelism, and inter-sequence parallelism are used in seeking performance improvement.
LOGAN takes advantage of the fact that each cell on the same anti-diagonal of the DP(dynamic programming) table is independent of the other and can be processed concurrently.
One GPU thread is spawned for each cell to overcome the limitation of available GPU threads and the DP table is split into segments. 
Furthermore, multiple pairs of sequences are processed in parallel by assigning each alignment to a GPU block. 

Experiments show that LOGAN improves the performance of up to 10.7$\times$ of BELLA, which gives us the rationale to combine LOGAN and ELBA together to solve the bottleneck of pairwise alignment.

\section{Method}
\label{sec:method}

We proposed 3 GPU schedulers compared with the baseline ELBA GPU scheduler.

\subsection{Baseline}
The vanilla GPU implementation of ELBA~\cite{Guidi2020ParallelSG} uses one MPI process and multiple GPUs.
The total work of one process is divided into batches of size $10,000$, and then each batch is divided into $c$ sub-batches. At each step, one sub-batch is fed to GPUs to perform the computation. The advantage of this approach is that since there is only one MPI process, it is free to use all of the GPU resources. There is no communication cost and racing condition. The downside is that it cannot benefit from using multiple MPI processes to accelerate pairwise alignment and other steps. Building on top of this baseline model, after introducing multiple MPI processes, we need to add a GPU scheduler to avoid more than one process using one GPU at the same time. We experimented with three scheduling algorithms to improve the pairwise alignment and the overall performance.

\subsection{\ota{} Scheduler}

\begin{figure}[htbp]
    \centering
    \includegraphics[width=0.9\linewidth]{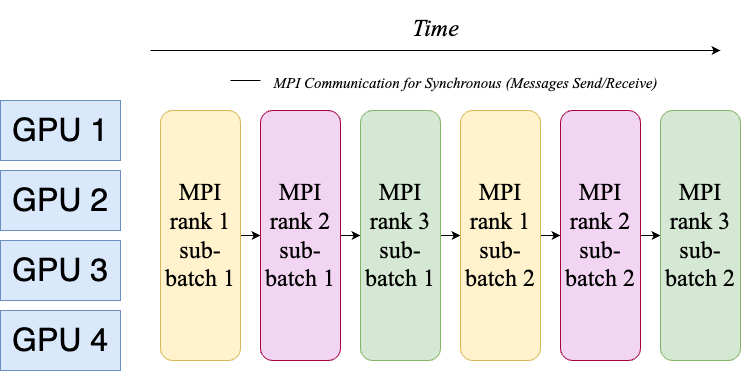}
    \caption{
    \textbf{\ota{} Scheduler Workflow}
    - Each MPI Process uses all GPUs, one sub-batch is scheduled each time
    }
    \label{fig:one2all}
\end{figure}

As showed in \cref{fig:one2all}, each MPI process uses all the GPU resources in a round-robin fashion.
One process schedules a sub-batch to all GPUs, and after computation is finished, it signals the next process that still has remaining work in the current round.
After receiving the completion message, the next process starts scheduling work to all GPUs.
The advantage of this algorithm is that it is easy to implement, the communication cost is low, and there is no load imbalance across GPUs.
The disadvantage is that even though each process has less work, still only one process is using GPU at a time.
Therefore, it does not fully exploit the benefit of having multiple processes.

We use \texttt{MPI\_Send} and \texttt{MPI\_Recv} to implement the signals because we want to create implicit barriers to force the MPI processes not to access GPUs at the same time.

\begin{algorithm*}[htbp]
\caption{\ota{} Scheduler} 
\label{alg:one2all}
\begin{algorithmic}[1]
    \State Set up $n$ MPI processes
    \State $rank \gets My\_MPI\_Rank$
    \State $completed \gets \{\}$
    \State $batches\_num \gets []$
    \For {$r=1,2,\ldots,n$}
        \If{$r==rank$}
        \State MPI\_Send(total\_batch,$[1,2,\ldots,rank-1,rank+1,\ldots,n]$)
        \Else
        \State MPI\_Recv(batches[r],[r])
        \EndIf
    \EndFor
    \For {$batch=1,2,\ldots, total\_batch$}
    \For {$iteration=1,2,\ldots,$}
            \If{$iteration==1$ and $batch==1$ and $rank==0$}
            
                \State Pairwise Alignment on GPU 
                \State MPI\_Send(True,[1])
            
            \Else
                \State $left \gets rank-1$\label{alg1:18}
                \While{$batch>batches[left]$ and $left \neq rank$}
                \State $left=left-1$
                \EndWhile
                \If{$left\neq rank$}
                \State MPI\_Recv(*,[left])
                \EndIf\label{alg1:24}
                \State Pairwise Alignment on GPU
                \State $right \gets rank+1$\label{alg1:26}
                \While{$batch>batches[right]$ and $right \neq rank$}
                \State $right=right+1$
                \EndWhile
                \State MPI\_Send(True,[right])\label{alg1:30}
            \EndIf
    \EndFor
    \EndFor
\end{algorithmic}
\end{algorithm*}

The corner cases such that rank-1 is smaller than 0 are taken care of by traversing in a ring-array fashion. Before executing, all MPI processes perform an all-to-all communication as specified from Line 5 to Line 11 to acknowledge how many batches each process has. This is important to avoid deadlocks and waiting for completed processes.

A detailed pseudo-code is presented at \cref{alg:one2all}.
We add this scheduler layer in the distributed batch runner and use LOGAN as a subroutine because MPI and CUDA can't be compiled together.
The master process will start first and signal the next process, and so on and so forth. The line \ref{alg1:18} to Line \ref{alg1:24} are used to create an implicit barrier. A completed process can be checked by comparing the batch number and the total batch number. We use a while loop to find the left-most uncompleted process. If such a process exists, the current process needs to wait for the signal from that process to avoid conflicts on GPU. Similarly, a process needs to find the right most uncompleted process to send a signal as specified from line \ref{alg1:26} to line \ref{alg1:30}.

\subsection{\oto{} Scheduler}

\begin{figure}[htbp]
    \centering
    \includegraphics[width=0.9\linewidth]{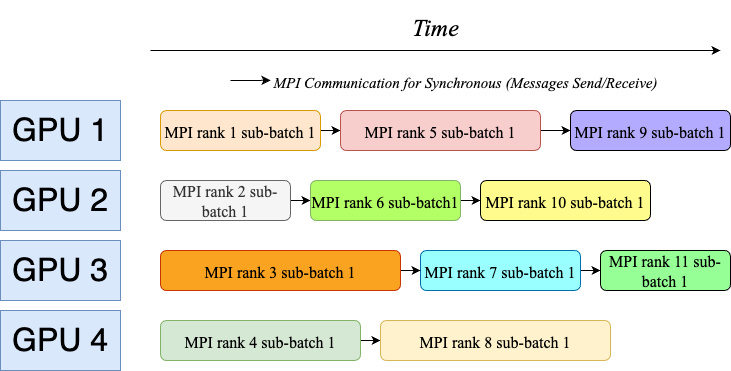}
    \caption{
    \textbf{\oto{} Scheduler Workflow}
    - Each MPI Process uses one GPU, one sub-batch is scheduled each time
    }
    \label{fig:one2one}
\end{figure}

As showed in \cref{fig:one2one}, processes are divided into pipelines where each pipeline has one dedicated GPU. Processes in the same pipeline schedule one sub-batch of work to its assigned GPU, and signals the next process in its pipeline after the job is finished. The scheduling inside one pipeline is round-robin. To be more specific, assume that there are $m$ GPUs, the $n$-th MPI process will be assigned to $n$ mod $m$ pipeline and perform computation on the $n$ mod $m$ GPU. Instead of traversing sequentially, we traverse with step size at the number of GPUs to create an implicit pipeline.

This approach achieves better parallelism because the number of processes that are active at a time is proportional to the number of GPUs. For the fixed sub-batch size, the performance of using only one GPU is not significantly worse than using all GPUs. However, the drawback of this approach is that each pipeline can have different total batches. This may cause some load imbalance on different GPUs. Also, if one GPU has higher computational power than others, it will become idle after it completes its own work.

\subsection{\ooto{} Scheduler}

\begin{figure}[htbp]
    \centering
    \includegraphics[width=0.9\linewidth]{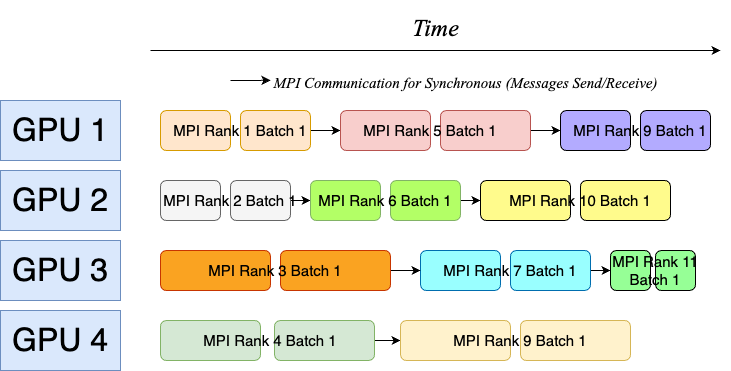}
    \caption{
    \textbf{\ooto{} Scheduler Workflow}
    - Each MPI process uses one GPU, one batch is scheduled each time, 
    }
    \label{fig:optone2one}
\end{figure}

To lower the communication cost of the previous approach, we came up with the idea of letting one process finish all the sub-batches in one batch before signaling the next process.
As showed in \cref{fig:optone2one}, we move the MPI communication from the iteration level to the batch level. The total communication cost should decrease proportionally to the number of sub-batches per batch. However, the GPU might be idling for a short period of time between the runs of sub-batches while the process is doing GPU-unrelated work.

\section{Results}
\label{sec:results}

\subsection{Experimental Setup}

We measure the total and alignment running time of the three schedulers proposed in \cref{sec:method}.
The experiments are divided into 2 groups: strong scaling with respect to the number of MPI processes and strong scaling with respect to the number of GPUs. 
The strong scaling experiments are performed on both \colismall{}(266 MB, 8,605 sequences) and \colilarge{} (929 MB, 91,394 sequences) datasets.

For reproducibility, all of the experiments are conducted on NERSC Perlmutter supercomputer cluster.
All runs are using either 1 or 2 nodes, each MPI process is assigned with 4 CPU cores.
Except for the computational resources, other hyperparameters and specifications are the same, namely

\begin{itemize}[nosep]
    \item --cpu-bind=cores, --gpu-bind=none
    \item -k (the k-mer length): 31
    \item -s (the k-mer stride): 1
    \item --alph (The alphabet to use): dna
    \item --ga (GPU-based x-drop alignment): 15
\end{itemize}

Codes for \colismall{} datasets use \texttt{LOWER\_KMER\_FREQ=20} and \texttt{UPPER\_KMER\_FREQ=30}, while codes for \colilarge{} datasets use \texttt{LOWER\_KMER\_FREQ=20} and \texttt{UPPER\_KMER\_FREQ=50}.

\subsection{Strong Scaling experiments w.r.t. number of MPI processes}

\begin{figure*}[htbp]
\begin{center}
\includegraphics[width=\linewidth]{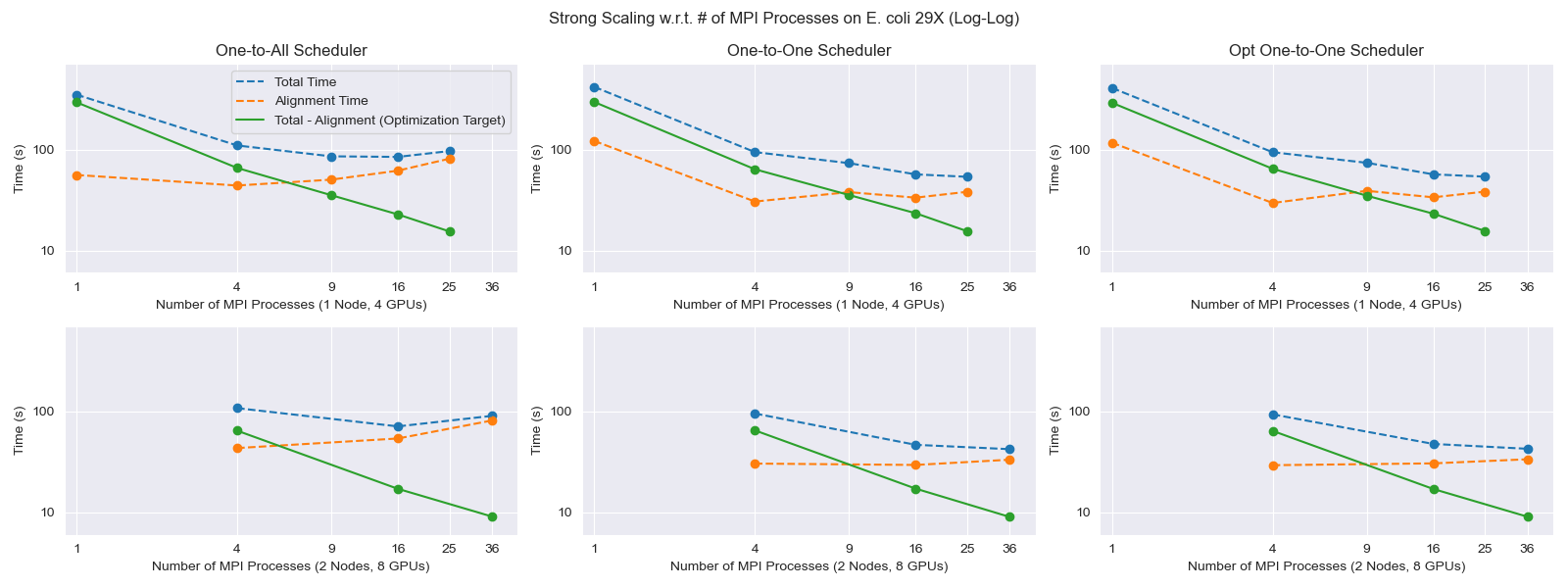}
\end{center}
\caption{
\textbf{Strong Scaling -- }
w.r.t. number of MPI processes on E. Coli 29x
}
\label{fig:strong_29x}
\end{figure*}

\begin{figure*}[htbp]
\begin{center}
\includegraphics[width=\linewidth]{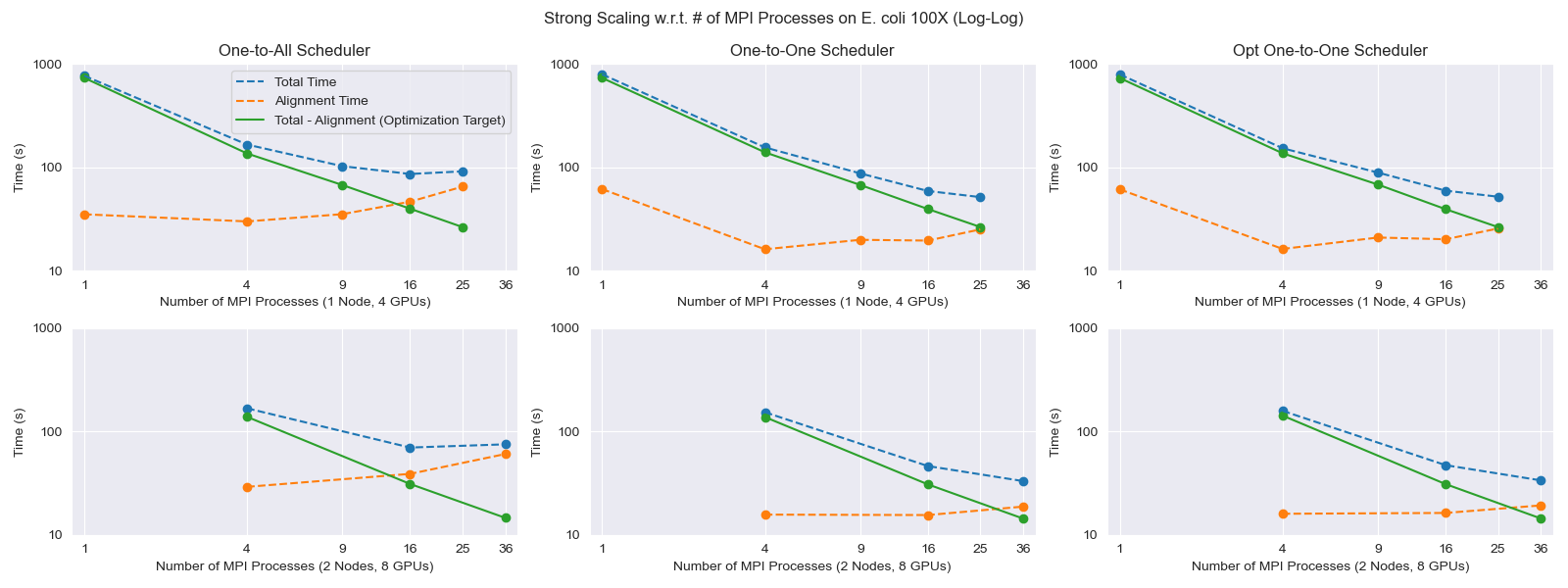}
\end{center}
\caption{
\textbf{Strong Scaling -- }
w.r.t. number of MPI Processes on E. Coli 100x
}
\label{fig:strong_mpi}
\end{figure*}

\cref{fig:strong_mpi} illustrates the strong scaling of our three approaches. 
We tested on \{1, 4, 9, 16, 25\} MPI processes and 4 GPUs. 
The setup with 1 MPI process is treated as the baseline because ELBA can only support 1 MPI process if CUDA is used in pairwise alignment. 
The best performance achieves approximately 10x speed up in total time with 4 MPI processes. 

The alignment time increases from 4 to 25 MPI processes, which is expected because the overhead of MPI communication increases linearly. However, the total runtime still decreases from 4 to 25 MPI processes because other steps also benefit from having more MPI processes, and this overweights the increasing communication overhead. One interesting observation is that the alignment time is faster for 4 and 9 MPI processes than only 1 MPI process. This provides evidence that our GPU scheduler benefits from MPI parallelism. Our implementation splits the data on the CPU concurrently before sending it to GPUs.

\subsection{Strong Scaling experiments w.r.t. number of GPUs}

\begin{figure*}[htbp]
\begin{center}
\includegraphics[width=\linewidth]{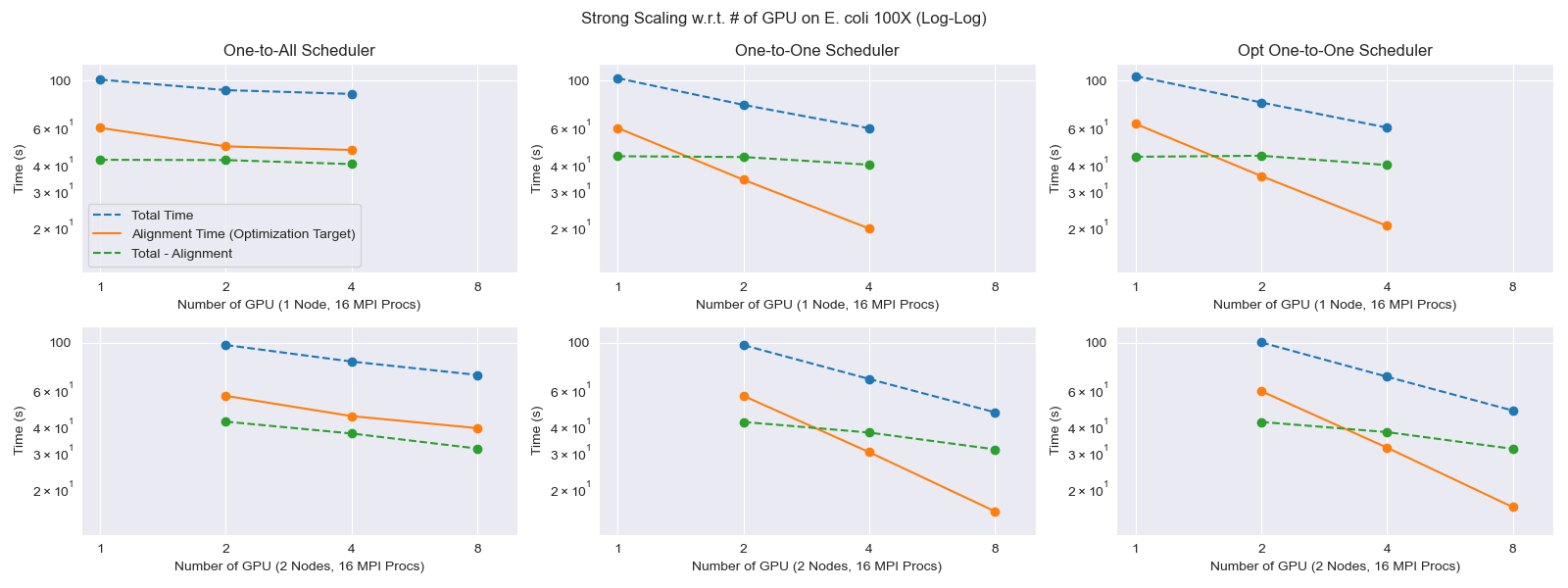}
\end{center}
\caption{
\textbf{Strong Scaling -- }
w.r.t. number of GPUs on E. Coli 100x
}
\label{fig:strong_gpu}
\end{figure*}

\cref{fig:strong_gpu} illustrates the strong scaling of our three approaches with respect to the number of GPUs.
We run three approaches on \{1, 2, 4\} GPUs with 16 MPI processes. 
Both alignment time and total time scales down, and (total alignment) time stays almost constant. This aligns with our hypothesis because having more GPUs means more MPI processes can make progress at the same time.
The difference between total time and alignment time scales down with an increasing number of MPI processes, and not with respect to the number of GPUs.
One interesting observation is that the alignment time of the one-to-one method is shorter than one-to-all. 
It is expected because multiple MPI processes can move data from CPU to GPU at the same time. 
Also, since we divide MPI processes into different pipelines, the communication overhead per pipeline is lower.

\subsection{Weak Scaling experiments w.r.t. number of MPI processes}

\begin{table*}[htbp]
\centering
\resizebox{\linewidth}{!}{ %
\begin{tabular}{ccccc|ccc}
\textbf{Dataset} &
  \textbf{\# Seq} &
  \textbf{Scheduler} &
  \textbf{\# MPI} &
  \textbf{\# GPU} &
  \textbf{\begin{tabular}[c]{@{}c@{}}Total\\ Time(s)\end{tabular}} &
  \textbf{\begin{tabular}[c]{@{}c@{}}Alignment\\ Time(s)\end{tabular}} &
  \textbf{\begin{tabular}[c]{@{}c@{}}Difference\\ Time(s)\end{tabular}} \\ \hline
E. coli 29X  & 8,605                & one2all      & 1  & 4 & 349.82 & 55.98  & 293.84                       \\
E. coli 100X & 91,394 ($\sim$10.6x) & one2all      & 16 & 4 & 86.43  & 46.57  & \textbf{39.85 ($\sim$7.37$\times$)} \\ \hline
E. coli 29X  & 8,605                & one2one      & 1  & 4 & 417.35 & 121.70 & 295.65                       \\
E. coli 100X & 91,394               & one2one      & 16 & 4 & 59.15  & 19.57  & \textbf{39.58 ($\sim$7.46$\times$)} \\ \hline
E. coli 29X  & 8,605                & opt\_one2one & 1  & 4 & 407.16 & 116.80 & 290.36                       \\
E. coli 100X & 91,394               & opt\_one2one & 16 & 4 & 59.59  & 20.16  & \textbf{39.44 ($\sim$7.36$\times$)} 
\end{tabular}
} %
\caption{
\textbf{Weak Scaling} -- w.r.t. number of MPI Processes
} %
\label{tab:weak}
\end{table*}

\cref{tab:weak} shows the weak scaling of our three approaches with respect to the number of MPI processes.
When the data size increases ~10.6 times, as the number of MPI processes increases 16 times, the difference in total time and alignment time only speed up ~7.4 times for all 3 schedulers, which shows a poor weak scaling efficiency constant.

More discussion will be covered in \cref{sec:result_big_small}.

\subsection{Comparison between small dataset and large dataset}
\label{sec:result_big_small}
After comparing the performance of \colilarge{} to the performance on \colismall{} as shown in \cref{fig:strong_29x}, we find that the strong scaling for the small dataset is worse than the large dataset. 
The overall run time increases as the number of MPI processes goes from 4 to 25.

However, referring to \cref{fig:strong_29x} and \cref{fig:strong_mpi}, surprisingly, neither the total runtime nor the alignment time of \colilarge{} experiments follows an $\mathcal{O}(n)$ or $\mathcal{O}(\log n)$ runtime complexity scale up, which makes the weak-scaling analysis above trivial.
The fundamental reason for it is that the computation workload needed by the LOGAN algorithm\cite{Zeni2020LOGANHG} (essentially Needleman-Wunsch or Smith-Waterman algorithm with X-Drop) is highly influenced by up-streaming K-mers related parameters like K-mers length, K-mers stride, lower and upper K-mers frequency. 

All the above experiments are conducted using Prof. Giulia Guidi's preset parameters where the lower kmer frequency is 20 and the upper kmer frequency is 50 for \colilarge{} datasets.

\section{Conclusions}
\label{sec:conclusions}

In this work, we introduce 3 GPU schedulers for ELBA to accommodate multiple MPI processes and multiple GPUs.
We show that 3 schedulers are able to significantly improve the performance of the baseline by conducting both strong and weak scaling experiments.
For the best performance implementation, the one-to-one scheduler achieves $\sim$7-8$\times$ speed-up using 25 MPI processes compared with the baseline vanilla ELBA GPU scheduler.

\section{Acknowledgement}
We are grateful to Professor Giulia Guidi for her tutorial on the background of ELBA and guidance on implementing this project.

{

    \bibliographystyle{IEEEtranN}
    \bibliography{main}
}

\end{document}